\documentclass[conference]{IEEEtran}
\IEEEoverridecommandlockouts
\usepackage{cite}
\usepackage{amsmath,amssymb,amsfonts}
\usepackage{algorithmic}
\usepackage{graphicx}
\usepackage{textcomp}
\usepackage{xcolor}
\usepackage{mathtools}

\usepackage{bbm}
\usepackage{mathrsfs}
\usepackage{float}
\usepackage[english]{babel}
\usepackage{setspace}
\usepackage{epstopdf}
\usepackage{textcomp}
\usepackage[linesnumbered,ruled,vlined]{algorithm2e}
\usepackage{dsfont}

\newtheorem{lemma*}{Lemma}

\renewcommand{\bar}{\overline}

\newcommand{\cm}[1]{\left(#1\right)}

\newcommand{\cX}{\mathcal{X}}

\def\BibTeX{{\rm B\kern-.05em{\sc i\kern-.025em b}\kern-.08em
    T\kern-.1667em\lower.7ex\hbox{E}\kern-.125emX}}
\begin{document}

\title{Majority is Not Required: A Rational Analysis of the Private Double-Spend Attack from a Sub-Majority Adversary\\
\thanks{Sriram Vishwanath is also an advisor to Dominant Strategies Inc.}
}
\author{\IEEEauthorblockN{Yanni Georghiades}
\IEEEauthorblockA{\textit{ECE Department} \\
\textit{UT Austin}\\
Austin, United States \\
yanni.georghiades@utexas.edu}
\and
\IEEEauthorblockN{Rajesh Mishra}
\IEEEauthorblockA{\textit{ECE Department} \\
\textit{UT  Austin}\\
Austin, United States \\
rajeshkmishra@utexas.edu}
\and
\IEEEauthorblockN{Karl Kreder}
\IEEEauthorblockA{\textit{CTO} \\
\textit{Dominant Strategies Inc.}\\
Austin, United States \\
karl@dominantstrategies.io}
\and
\IEEEauthorblockN{}
\and
\IEEEauthorblockN{Sriram Vishwanath}
\IEEEauthorblockA{\textit{ECE Department} \\
\textit{UT Austin}\\
Austin, United States \\
sriram@utexas.edu}

% \and
% \and
% \and
}

\maketitle

\begin{abstract}
We study the incentives behind double-spend attacks on Nakamoto-style Proof-of-Work cryptocurrencies. 
In these systems, miners are allowed to choose which transactions to reference with their block, and a common strategy for selecting transactions is to simply choose those with the highest fees. 
This can be problematic if these transactions originate from an adversary with substantial (but less than 50\%) computational power, as high-value transactions can present an incentive for a rational adversary to attempt a double-spend attack if they expect to profit.
The most common mechanism for deterring double-spend attacks is for the recipients of large transactions to wait for additional block confirmations (i.e., to increase the attack cost).
We argue that this defense mechanism is not satisfactory, as the security of the system is contingent on the actions of its \textit{users}. 
Instead, we propose that defending against double-spend attacks should be the responsibility of the \textit{miners}; specifically, miners should limit the amount of transaction value they include in a block (i.e., reduce the attack reward). 

To this end, we model cryptocurrency mining as a mean-field game in which we augment the standard mining reward function to simulate the presence of a rational, double-spending adversary.
We design and implement an algorithm which characterizes the behavior of miners at equilibrium, and we show that miners who use the adversary-aware reward function accumulate more wealth than those who do not. 
We show that the optimal strategy for honest miners is to limit the amount of value transferred by each block such that the adversary's expected profit is 0.
Additionally, we examine Bitcoin's resilience to double-spend attacks under our model.
Assuming a 6 block confirmation time, we find that an attacker with at least 25\% of the network mining power can expect to profit from a double-spend attack. 
\end{abstract}

\section{Introduction}
\label{sec:intro}
This paper characterizes the behavior of cryptocurrency mining in the presence of a rational adversary who is actively  attempting adouble-spend attack. 
We focus specifically on Nakamoto-style Proof-of-Work cryptocurrencies, in which miners compete to add blocks to the blockchain.
A miner who successfully appends a block to the blockchain wins rewards in the form of a constant block reward and a set of transaction fees that correspond to the transactions the miner references in the block. 

Commonly, miners seek to maximize the transaction fees they  receive by greedily selecting the transactions with the highest fees.
However, this strategy is problematic in the presence of a rational adversary with the means to execute a double-spend attack, as the amount of economic value being transferred by a transaction (henceforth referred to as the ``transaction value") is directly related to the adversary's expected profit in attempting such an attack. 
To combat this problem, we present a mining reward function which accounts for the transaction value in the block and the impact it has on the adversary's expected profit. 
Using this new reward function, we model the mining problem as a dynamic mean-field game in which agents choose the amount of transaction value to include in their block at each time step.
We provide an algorithm which solves for equilibrium strategies in this game in order to compare the behavior and performance of miners utilizing the standard reward function (and greedily selecting the highest fee transactions) with that of miners utilizing our adversary-aware reward function. 

\subsection{Motivation}
% There is a popular belief that a Proof-of-Work miner with a significant fraction of the total network mining power will refrain from attempting double-spend attacks because public knowledge of such an attack would devalue their investment. 
% % this sentence is clunky ^
% This belief is likely reasonable for some miners, as dedicated Proof-of-Work mining operations require a significant up-front investment of capital to purchase equipment and facilities.
% However, it is also reasonable for a rational adversary to determine that such attacks are indeed profitable and proceed with them.
% For this reason, it is worthwhile to understand the security limitations of each system so that one might subsequently design systems that can defend against such attacks. 
% This is especially prescient for newer and smaller cryptocurrencies which may have relatively low hash power and are therefore easier targets.
% In this work, we show that by modifying the transaction selection mechanism to account for the value of a transaction and imposing a transaction fee which increases with the transaction value, we can render double-spend attacks unprofitable against rational adversaries while still enabling miners to follow incentive-compatible strategies.

Studying the incentives a rational miner has to attempt a double-spend attack and designing protocols which are resilient to such an attack is critical to the long-term utility of Proof-of-Work cryptocurrencies. 
However, some would refute such a position, claiming that no rational adversary would be incentivized to execute such an attack because either 1) Bitcoin receives so much mining power that it is not feasible for an adversary to succeed in an attack; or 2) any such adversary must have a significant investment in the cryptocurrency, and attacking the currency would devalue their investment to such an extent that any attack would be ultimately unprofitable. 

The first argument is specific to Bitcoin and does not apply to all Proof-of-Work cryptocurrencies, particularly newer and smaller cryptocurrencies which may be easier targets due to their relatively low network mining power. 
The second argument, although seemingly persuasive, makes implicit assumptions on the adversary's financial model which may not hold in all cases.
For example, \cite{bonneau2016buy} argues that it would be straightforward for attackers to temporarily borrow mining power from other miners through bribery, vastly reducing the investment required to prepare for the attack.  
Even simpler, adversaries may rent mining power directly from a third party provider such as nicehash.com; indeed, in 2020 this strategy was used to great effect in attacks against Bitcoin Gold  and Ethereum Classic, resulting in the theft of assets valued in the millions of dollars \cite{btg2020, etc2020}. 
Moreover, a decline in currency value could even be advantageous to a prepared adversary who holds ``short"-positions on the value of that currency.
Overall, relying on an adversary's aversion to price volatility to refrain from attacks is not a satisfactory solution, and more careful analysis is needed to understand both the impact and the associated mitigation strategies of such attacks.

\subsection{The Relationship Between Transaction Values and Transaction Fees}
In  this work, we show that the cost an adversary incurs in attempting a double-spend attack increases with the total network mining power. 
We also observe that the total network mining power increases as mining rewards increase because miners naturally seek to claim a share of the extra rewards. 
In order to incentivize sufficient mining power to deter attacks, high-value transactions (which are more attractive targets) should pay higher fees than low-value transactions.  
Intuitively, high-value transactions should pay for the security required to safeguard them.

However, in typical cryptocurrencies, transaction fees are unrelated to transaction value, as fees are typically market-driven based on the demand for transaction settlement. 
%Transaction values are largely ignored despite the significant impact they have on system security.
To remedy this issue, we propose a reward model in which transaction fees increase monotonically with transaction values. 
Specifically, in this work the transaction fee is a small percentage of the transaction value.
Although such a  fee model may or may not be optimal, it serves as a good starting point for analysis. 
We believe the study of transaction fees as they relate to transaction values is  critical to the long-term security of Proof-of-Work cryptocurrencies, but we defer further analysis of this topic to future work. 

% For such an adversary, the transaction value is the reward an adversary expects to receive for a successful double-spend attack.
% In order to discourage such an adversary from attacking, higher transaction values must be associated with a higher cost to attack. 
% It is clear from Equation \ref{eq:adv_cost} that the adversary's attack cost increases with the total network mining power.
% It is also safe to assume that at a competitive equilibrium in any cryptocurrency mining market, total network mining power increases with total mining rewards (which can come in the form of transaction fees).
% In other words, the adversary's attack reward increases with transaction value, and the adversary's attack cost increases with transaction fees. 
% We therefore argue that, in general, transaction fees should increase monotonically with transaction value. 
 
\subsection{Mean-Field Games}

Mean-field games (MFGs) were introduced in~\cite{Huang2006, Lasry2007} as an efficient mechanism to characterize the interdependent behavior of agents in a  system with a large number of participants. 
They involve aggregating the behavior of all agents into an associated \emph{mean-field state} that is then used by individual agents to make rational decisions. 
Cryptocurrency mining is a game in which the decisions of each miner are determined by the observable and expected behavior of each other miner.
Although miners may not know the precise mining power of their peers, the total mining power employed by the network to mine each block can be approximated by observing block times. 
Consequently, we model the average mining power employed by each miner as a mean-field state, meaning miners know the average mining power contributed by the other agents but not any specific miner's contribution.
In Section \ref{sec:model}, we present a mean-field game model for cryptocurrency mining in the presence of a rational adversary. 

In Section \ref{sec:implementation}, we provide an algorithm which simulates our game and can be used to determine the MFG equilibrium policies for miners.
We implement this algorithm in Python3 and use it to compare the mining performance for miners that do account for the adversary in their reward function and those that do not.
We show in Section \ref{sec:perf} that if miners use an adversary-aware reward function, the optimal equilibrium policies render double-spend attacks unprofitable for the adversary. 
On the other hand, if miners ignore the threat of adversarial attack and select high transaction values, they experience a significant reduction in profitability when the adversary does attack.

Finally, in Section \ref{sec:bitcoin} we simulate the security of Bitcoin under our model using coarse estimations on current Bitcoin network parameters. 
We find that under a 6 block confirmation time, an adversary with at least 25\% of the network mining power can expect to profit from a double-spend attack. 
However, we show that implementing a small percentage-based transaction fee (between 1.25-1.5\%) would result in additional mining power to secure current levels of transaction value. 

\subsection{Our Contributions}

In summary, the key contributions of this paper are:

\begin{itemize}
    \item A framework that models the costs and rewards of a rational adversary who has the ability to execute double-spend attacks.
    \item A characterization of the mining game in the presence of a rational adversary as a mean-field game. 
    \item Detailed pseudo-code for an algorithm which solves for the equilibrium strategies of agents playing this game. 
    \item Demonstration of the optimal strategies for miners using an adversary-aware reward function, under which a double-spend attack is always unprofitable for a rational adversary.
    \item Analysis of the Bitcoin's security under our model, in which we show that an adversary with at least 25\% of the network mining power could expect to profit from a double-spend attack.
\end{itemize}

\section{Related Work}
\label{sec:related}
There is a considerable body of work in understanding blockchains using mechanism design and game theory.
For a game theory perspective, we have a variety of different approaches \cite{kiayias2016blockchain, ewerhart2020finite,singh2020game,yuan2020framework,liu2019survey,goren2019mind,sun2020games,min2019security,altman2020blockchain,wang2019pool,ferreira2021proof}. 
Orthogonal to our analysis of transaction fees, there is a rich body of literature on using mechanism design to implement incentive compatible and collusion-resistant transaction fee mechanisms \cite{lavi2022redesigning, chihincentive, roughgarden2021transaction,roughgarden2020transaction, chung2023foundations}. 
Below we discuss work more closely related to ours. 

\subsection{Game Theoretic Analysis of Double-Spend Attacks}
% We extend a brief line of work analyzing double-spend attacks in terms of the adversary's likelihood of success and expected profitability. 
In the original Bitcoin whitepaper \cite{nakamoto2008bitcoin}, Nakamoto first estimates the probability of a successful double-spend attack with an infinite time horizon. 
Nakamoto models the number of blocks the attacker can mine in a given time period as a Poisson distribution and then applies a gambler's ruin argument to find the probability that the attacker can ever catch up to or overtake the honest chain. 
Importantly, Nakamoto only examines the probability of success and does not model the economic cost of the attack.
Rosenfeld \cite{rosenfeld2014analysis} later enhances the accuracy Nakamoto's analysis by modeling the number of blocks mined by the attacker with a negative binomial distribution.
Rosenfeld briefly examines the economics of the double-spend attack, suggesting heuristics that merchants can use to determine whether or not it is safe to accept a payment.

Bissias \textit{et al.} \cite{bissias2016analysis} and Sompolinsky \textit{et al.} \cite{sompolinsky2016bitcoin} further examine double-spend attacks from an economic perspective.
They model the profitability of the double-spend attack, and they also consider an adversary that can leverage additional attack vectors to enhance the effectiveness of their attack. 
Recently, Jang \textit{et al.} \cite{jang2020profitable} propose a more finely-grained model of the attacker's expected profits using a Markov decision process, allowing the attacker to re-evaluate at each time step whether or not it is profitable to continue the attack.

While most of these works examine the double-spend attack in terms of the profitability of the adversary, the only defense mechanisms suggested are to be carried out by the merchant. 
The merchant can either decline large transactions or require extra block confirmations before the goods exchange hands. 
In contrast, we consider a scenario in which the onus of defense is placed on the miners, allowing the merchant to remain agnostic to the inner workings of the blockchain. 
This aligns with the incentives of the ``honest" miners (which we still presume to be economically motivated), as double-spend attacks are costly for miners who lose out on block rewards as a result of the attack. 
We show that it is in the miners' best interests to implement our suggested defense of limiting the transaction value made vulnerable to attack. 
In so doing, we absolve merchants from the responsibility of defending the system against attacks and place that responsibility in the hands of the miners. 

\subsection{The Rational Protocol Design Framework}
A separate line of work studying this problem has its foundations in the Rational Protocol Design (RPD) framework proposed by Garay \textit{et al.} \cite{garay2013rational} as a means to analyze cryptographic protocols under the assumption that participants are rational rather than honest or corrupt. 
The RPD framework considers a two-party game between the protocol designer and an attacker in which the protocol designer specifies a protocol and the attacker specifies a polynomial-time attack strategy to subvert the protocol. 
The attacker gains utility by violating security guarantees of the protocol but must pay to corrupt protocol participants, and the protocol is secure if the adversary cannot propose any attack strategy which yields positive utility.

Badertscher \textit{et al.} \cite{badertscher2018but} later extend the RPD framework in order to study Bitcoin from a rational perspective. 
They find that, even if the majority of miners are not assumed to be honest, Bitcoin is incentive compatible (meaning all parties follow the protocol) if the transaction fees available to miners do not vary significantly between blocks. 
This result does not apply to our model, as their utility functions explicitly ignore the transaction value (which is a focal point in our analysis). 

More recently, Badertscher \textit{et al.} \cite{badertscher2021rational} expand on their previous work to study the problem of a 51\% double-spend attack. 
To this end, they devise a utility function which captures the incentives of a double-spending attacker with the ability to take over a majority share of the network mining power for a period of time. 
Using this utility function, they characterize a range of attack payoffs that the attacker would have to receive in order for the attack to be profitable. 
Finally, they show that by increasing the number of block confirmations required for a transaction to be considered finalized, the incentive for a rational adversary to attempt a 51\% double-spend attack can be eliminated. 
Out of all works discussed, our model is most similar to theirs in that both analyze the incentives and costs for an attacker to attempt a private double-spend attack and both suggest a mechanism by which that attack can be mitigated. 
However, their model assumes that the attacker is able to acquire a majority share of mining power for a period of time long enough to guarantee that the attack is successful, whereas we consider the scenario in which the attacker does not control a majority share and is not guaranteed to succeed. 
The structure of our games are also dramatically different, as we define a dynamic mean-field game in which each round corresponds to a block being mined, while they consider a Stackelberg game of two rounds (the protocol designer proposes a protocol and then the adversary proposes an attack). 
As a result, our reward functions and analysis methods vary significantly from theirs, where we define exact reward functions to simulate the attacker's decision on whether or not to attack and they provide bounds on their utility functions under which the attacker is incentivized to attack.
Finally, our proposed defense mechanism is a limitation on the value which can be transacted in a given block, while theirs is a modification to the number of block confirmations required for a transaction to be finalized. 
Each of these mechanisms has practical implications which might make one more appealing than the other depending on application requirements, but we argue that entrusting system security to the actions of its users is not a satisfactory solution.

% \subsection{Transaction Fee Mechanism Design}
% A related area in which mechanism design has been applied to cryptocurrencies is in the design of transaction fees 

\subsection{Mean-Field Games}

Mean-field games have proven effective in modeling large population games with interdependent agents. 
Recently, \cite{li2019mean} models the cryptocurrency mining problem as a mean-field game and derives equilibrium policies in terms of the mining power and wealth of the miners. 
They provide both an analytical framework and numerical algorithm to derive equilibrium policies for miners with infinite and finite wealth, respectively. 
We build upon their model to introduce an adversary and study how the behavior of miners is affected as a result. 
Our proposed algorithm is also derived from the work by authors in~\cite{mishra2020model} where they solve for equilibrium policies in mean-field games through a sequential decomposition algorithm.

Separately, Bertucci \textit{et al.} \cite{bertucci2020mean} study Bitcoin mining as a mean-field game.
Their analysis differs from that of \cite{li2019mean} in that it focuses on hashrate stability in response to changing environmental variables such as technological progress, electricity costs, and currency conversion rates.
While interesting, we do not consider such variables in our model.

\section{Model}
\label{sec:model}
We model the mining game as a finite, sequential mean-field game  of $m$ agents. 
Our model includes a static adversary that controls a $\beta$-fraction of the network hash power at all times.
We refer to non-adversarial miners as either ``honest miners" or ``miners." 
Honest miners behave rationally within their action space, but that action space does not include the ability to attempt a double-spend attack of their own. 
The adversary, on the other hand, does not participate in mining unless actively executing a double-spend attack\footnote{see Section \ref{sec:conclusions} for additional discussion on this assumption}.
In other words, there is a clear delineation in allowable actions between the adversary (who does not mine unless attacking) and the honest miners (who do not attack). 

At each time step, we define the mean-field variable $\bar{\alpha}_t$ as the average mining power used by all agents in that time step.  
Miners use $\bar{\alpha}_t$ in their decision-making process as an approximation of the competition they expect from the other miners.

For ease of analysis, we assume a synchronous network model for mining with no communication delays. 
Mining is performed in a series of rounds, where there is exactly 1 block mined per round.
Miners have limited mining resources (referred to as ``wealth") which carry forward as state between time steps.
The amount of wealth a miner has at a particular round limits the amount of mining power they can contribute to that block, and after each round all wealths are updated in accordance with the actions taken and the expected rewards they would receive as a result.
In a given round, the adversary (if they are attempting an attack) mines the block with probability $\beta$ and some honest miner mines the block with probability $1 - \beta$.

\subsection{The Adversary}
We consider a rational, myopic, risk-neutral adversary whose objective is to maximize their expected reward at each time step in the game without consideration of future rewards.
The adversary makes only one choice at each time step: they can attempt a private double-spend attack against the system or do nothing, and they only choose to attack if their expected reward from attempting the attack is greater than their expected cost\footnote{In order to smooth the reward function, in our simulation there is technically a negligible but non-zero probability that the adversary attacks in spite of a slightly negative expected profit. This does not impact the results.}.

\subsubsection{The Double-Spend Attack}
For our purposes, a private double-spend attack is performed as follows:

\begin{itemize}
    \item The adversary submits a transaction to the blockchain in payment for some asset X which is external to the system (for example, X could be a physical asset such as a yacht, or X could be a digital asset such as a token from another cryptocurrency).
    Privately, the adversary begins mining an alternate private chain in which they instead send that payment to an address that they control.
    \item After $k$ confirmation blocks are published to the public blockchain, the adversary receives the asset X.
    \item The adversary publishes their private chain (if it is longer than the public chain), retaining X and also regaining control of the funds they originally used to purchase X.
\end{itemize}

The adversary is successful in their attack if and only if they are able to mine $k+1$ blocks before the honest miners are able to mine $k+1$ blocks. 
In other words, the adversary makes a single decision in each time step of whether or not to attempt an attack, and if they do attempt the attack then they carry it out precisely until either the adversary or the honest miners reach $k+1$ blocks.
Although of significant interest, we defer the study of more complex adversarial strategies to future work. 

In terms of the amount of transaction value made vulnerable to attack, we consider a worst-case scenario in which \textit{all} of the transactions contained within a block can be simultaneously double-spent by a single adversary. 
While this may not always be achievable, a sophisticated adversary could ostensibly execute such an attack with foreknowledge of the honest miners' transaction selection strategies.

The adversary's decision on whether or not to attack is represented by the function $A$, which evaluates to 1 if the expected profit of an attack is positive and 0 otherwise.

\begin{equation}
\label{eq:adv_decision}
        A(T, \bar{\alpha}) =
  \begin{cases}
    1 & \text{if $R_{adv}(T, \bar{\alpha}) > 0$} \\
    0 & \text{else} \\
  \end{cases}
    \end{equation}

\begin{figure}[htp]
    \centering
    \includegraphics[width=6cm]{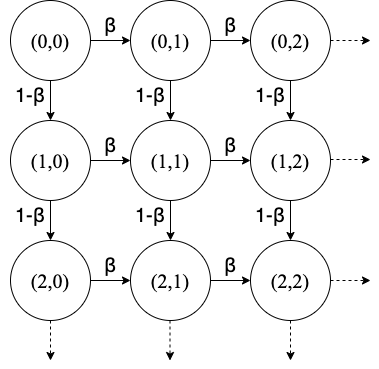}
    \caption{The 2-D Markov chain used to approximate the probability that the adversary can succeed in a double-spend attack and the expected number of blocks that the attack takes. States are represented by the tuple $(B_H, B_A)$, where $B_H$ is the number of blocks mined by the honest miners since the attack began and $B_A$ is the number of blocks mined by the adversary. $B_A$ is incremented with probability $\beta$, and $B_H$ is incremented with probability $(1 - \beta)$.}
    \label{fig:markov}
\end{figure}

In order to characterize the adversary's expected attack cost and probability of success, $C(\bar{\alpha}, \beta)$ and $P(\beta)$, we model the double-spend attack at a particular block as the 2-D Markov chain depicted in Figure \ref{fig:markov}.
Each state is a tuple $(B_H, B_A)$, where $B_H$ is the number of blocks the honest agents have mined and $B_A$ is the number of blocks the adversary has mined. 
From an initial state $(b_H, b_A)$, the transition to $(b_H, b_A + 1)$ occurs with probability $\beta$ and the transition to $(b_H + 1, b_A)$ occurs with probability $1 - \beta$. 
% The adversary succeeds in their attack upon any transition from a state $(b_H, k)$ to $(b_H, k+1)$, where $b_H \leq k$, and fails otherwise.
Then $P(\beta)$ is defined simply as 

\begin{equation}
    P(\beta) = \sum_{b_H = 0}^k \beta Pr[(b_H, k)],
\end{equation}

where $\beta Pr[(b_H, k)]$ is the probability of reaching state $(b_H, k)$ and then transitioning into state $(b_H, k+1)$.
In other words, $P(\beta)$ is the probability that an adversary with a $\beta$ fraction of the network mining power mines $k+1$ blocks while the honest agents have mined at most $k$ blocks.
% We model $P(\beta)$ in this way so that the expected number of blocks required to attack and the expected cost of attack can be precisely defined.

We define $C(\bar{\alpha}, \beta)$ similarly as 

\begin{equation}
\label{eq:adv_cost}
\begin{split}
    C(\bar{\alpha}, \beta) =&\beta  m \bar{\alpha} c \big((2k + 1) (1-P(\beta)) + \\ &\sum_{b_H=0}^{k}  \beta Pr[(b_H, k)](k+1+b_H) \big),
\end{split}
\end{equation}

% \textcolor{red}{should probably break this into two equations for readability}

where $\beta  m \bar{\alpha} c$ is the mining cost the adversary pays per time step ($m \bar{\alpha}$ is the total network mining power and the adversary controls a $\beta$ fraction of it), and the remainder of the expression is the expected number of blocks  the adversary attempts to mine as they execute the attack. 

\subsubsection{The Adversary's Reward Function}
Recall that $T$ is the amount of value being transacted within a particular block.
The adversary's expected reward when attempting a double-spend attack is

\begin{equation}
\label{eq:adv_reward}
    R_{adv}(T, \bar{\alpha}) = P(\beta) \big((k+1)b + T + f(T)\big)- C(\bar{\alpha}, \beta),
\end{equation}

where $(k+1)b$ corresponds to the $k+1$ block rewards the adversary wins, and $T$ and $f(T)$ are the transaction value and transaction fees, respectively, that they recuperate as a result of the double-spend attack.
% \textcolor{red}{redo this}

Note that because the adversary has a fixed strategy and does not carry state forward throughout the game (they are assumed to have a $\beta$ fraction of the mining power at their disposal at all times), they can more or less be thought of as a part of the environment.

\subsection{Honest Miners}
Miners are assumed to be rational\footnote{Honest miners are rational in the sense that they select the best action within their action space. We leave examination of a model in which miners can ``switch sides" and attempt a double-spend attack of their own to future work.}, risk-neutral, and forward-thinking. 
At each time step, miners select the actions that maximize the sum of their expected reward at that time step and the future rewards they expect to receive over the remainder of the game.

Honest miners choose two actions: (1) the amount of mining power $\alpha$ that they contribute towards mining a block, and (2) the total amount of transaction value $T$ that they choose to reference in the block, where the transaction fees they will receive upon mining a block are determined by a fee function $f(T)$.
We abstract $T$ and $f(T)$ in this way because attempting to simulate transaction pools and more complex transaction selection mechanisms is needlessly cumbersome. 
Notably, we constrain the action space of each miner such that the miners cannot select an $\alpha$ that exceeds their total wealth, as no miner can have infinite wealth.
We define a system parameter $T_{max}$ to be the maximum total transaction value that can be referenced by a block, and we assume that this value is always available to miners (i.e., the amount of available transaction value does not change between rounds).

\subsubsection{Reward Function}

We define two reward functions for the honest miners: the ``naive" reward in which miners are not aware of or not responsive to the threat of adversarial attack, and the actual reward which does account for the threat of attack. 
All reward functions in this work are expectations taken over the probability of successfully mining a block. 

We first define the probability that the miner wins a block and receives a reward as 

\begin{equation}
    \Gamma_{naive}(\bar{\alpha}, \alpha, T) = \frac{\alpha}{\alpha+m\bar{\alpha}}.
\end{equation}

Recall that $\alpha$ is the amount of mining power a miner decides to contribute to mining a particular round, so $\alpha + m\bar{\alpha}$ is the miner's best estimate on the total amount of mining power contributed by the rest of the network. 

Then the naive reward for honest miners is 

\begin{equation}
    R_{naive}\cm{\bar{\alpha}, \alpha, T} = \Gamma_{naive}(\bar{\alpha}, \alpha, T) \big(b + f(T)\big) - \alpha c,
\end{equation}

where $b$ is the constant block reward and $c$ is the mining cost per unit of mining power.
Combined with transaction fees, $b + f(T)$ is the total reward the miner receives for winning the block, and $\alpha c$ is the total cost to mine in that round. 
Recall that $f(T)$ increases monotonically with $T$, meaning $R_{naive}$ also increases with $T$. 
The best choice of $T$ under this reward function is therefore always $T_{max}$, which intuitively aligns with the current status quo strategy (greedily select the transactions with the highest fees).

On the other hand, if we consider the threat of adversarial attack, the probability that the miner receives the reward is instead 

\begin{equation}
    \Gamma(\bar{\alpha}, \alpha, T) = \big(1 - P(\beta) A(\bar{\alpha}, T)\big) \frac{\alpha}{\alpha+m\bar{\alpha}},
\end{equation}

Intuitively, if the adversary attempts an attack, then $\big(1 - P(\beta) A(\bar{\alpha}, T)\big)$ can be simplified to $\big(1 - P(\beta)\big)$, which is the probability that the attack fails. 
If the attack were to succeed, the honest miner would receive no rewards for the block.  

The adversary-aware reward function is then

\begin{equation}
\label{eq:honest_reward}
    R\cm{\bar{\alpha}, \alpha, T} =  \Gamma(\bar{\alpha}, \alpha, T) \big(b + f(T)\big) - \alpha c,
\end{equation}

In contrast with $R_{naive}$, $R$ does not necessarily increase with $T$, because a higher transaction value increases the likelihood of adversarial attack through $\Gamma$.

\subsection{Transaction Fees}
Our transaction fee function $f(T)$ is monotonically increasing in $T$, i.e., transactions with higher transaction values must pay out higher fees.
We model $f(T)$ in this way for two reasons. 

The first is that, as mentioned in Section \ref{sec:intro}, the incentive for the adversary to attack increases with transaction value.
Higher fees can be used to offset the risk imposed by higher value transactions because they serve as incentives for miners to contribute additional mining power to each block (thereby increasing the adversary's cost to attack).

The second reason is that defending against double-spend attacks is trivial if high value transactions do not pay the highest fees. 
In this case, there would be no reason for an honest miner to include a high value transaction in their block, as they could receive the same or greater rewards without risk by including only low value transactions. 
By defining $f(T)$ to be increasing in $T$, we impose a tradeoff between the rewards received for mining and the risk of adversarial attack, which allows us to demonstrate that miners willingly choose to receive lower fees in order to prevent the adversary from attacking. 
In practice, this tradeoff would likely exist in some form, as the adversary would construct transactions which pay high fees to ensure inclusion in the block. 

In our tests, we define $f(T)$ as 

\begin{equation}
    f(T) = \lambda T,
\end{equation}

where $\lambda$ is typically 0.01 in our experiments (i.e., a 1\% transaction fee). 
While there are many different fee functions that we could have chosen, a ``percentage-based" fee is both computationally convenient and easy to understand.
% Requiring that $f(T)$ is \textit{at least} $\lambda T$ guarantees that miners are compensated for taking on additional risk of adversarial attack.
% Additionally, and perhaps more importantly, limiting $f(T)$ to be \textit{at most} $\lambda T$  inhibits the ability for the adversary to ``bribe" miners into selecting high transaction values by offering extremely high fees. 
We defer further exploration of alternate transaction fee functions to future work.

\section{Solving the Mining Game}
\label{sec:implementation}

\begin{algorithm*}
    \label{alg:Evaluation_alpha}
    \DontPrintSemicolon
    \SetAlgoLined
    \KwIn
    {\\
        m: Number of miners\\
        c: Mining cost per unit of mining power\\
        b: Block reward per block\\
        $\gamma$: Momentum parameter\\
        $\tau$: The number of time steps\\
        $\bar{\alpha}_{0}$: Initial mean mining power\\
        $w_{0}$: Initial mean wealth \\
        % $V_{\tau +1} = 0$ The value function at time $\tau +1$ is 0 for all $x$ \\
        % $f(T): \mathbb{R} \rightarrow \mathbb{R}$ : Fee function
    }
    \KwOut{$\alpha_{N}, T_{N}$}
    \For{each $n$ until $N$}
    {

    \For{$t = \tau,\tau-1,\ldots, 0$}
    {
        % $R\cm{\bar{\alpha}, \alpha, T} = (1 - P(\beta) A(\bar{\alpha}, T)) (b + f(T))\frac{\alpha}{\alpha+m\bar{\alpha}} - \alpha c$\;
        $\forall x \in \cX$  $\alpha_{t,n}^\star(x), T_{t,n}^\star(x) = \arg \max_{\alpha,T} R\cm{\bar{\alpha}_{t,n}, \alpha, T} + \Gamma(\bar{\alpha}_{t,n}, \alpha, T) V_{t+1,n}\cm{x + R\cm{\bar{\alpha}_{t,n}, \alpha, T}} + \big(1-\Gamma(\bar{\alpha}_{t,n}, \alpha, T)\big) V_{t+1,n}\cm{x - \alpha c} $\;
        $\forall x \in \cX$  $V_{t,n}\cm{x} = R\cm{\bar{\alpha}_{t,n}, \alpha^\star(x), T^\star(x)} + \Gamma(\bar{\alpha}_{t,n}, \alpha^\star(x), T^\star(x))V_{t+1,n}\cm{x + R\cm{\bar{\alpha}_{t,n}, \alpha^\star(x), T^\star(x)}} + \big(1-\Gamma(\bar{\alpha}_{t,n}, \alpha^\star(x), T^\star(x))\big) V_{t+1,n}\cm{x - \alpha c}$\;
        
        $\forall x \in \cX$ $\alpha_{t,n}\cm{x} \leftarrow \alpha^\star(x)$\;
        $\forall x \in \cX$ $ T_{t,n}\cm{x} \leftarrow T^\star(x)$\;
    }
    \For{$t = 0,\ldots, \tau$}
    {
        % $\forall x \in \cX$  $w_{t+1,n}\cm{x} = w_{t,n}\cm{x+\alpha_{t,n}\cm{x} c} (1 - \frac{\alpha_{t,n}\cm{x}}{\alpha_{t,n}\cm{x}+m\bar{\alpha}_{t,n}} ) + w_{t,n}\cm{x+\alpha c - b}\frac{\alpha_{t,n}\cm{x}}{\alpha_{t,n}\cm{x}+m\bar{\alpha}_{t,n}} $\;
        
        $w_{t+1,n} \leftarrow $ \textit{zeros}\;
        $\forall x \in \cX$\;
        \hskip0.5em $x_{win} \leftarrow x + R\cm{\bar{\alpha}_{t,n}, \alpha_{t,n}\cm{x}, T_{t,n}\cm{x}}$\;
        \hskip0.5em $x_{lose} \leftarrow x -\alpha_{t,n}c $\;
        \hskip0.5em $w_{t+1,n}\cm{x_{win}} += \Gamma(\bar{\alpha}_{t,n}, \alpha_{t,n}\cm{x}, T_{t,n}\cm{x}) w_{t,n}\cm{x} $\;
        \hskip0.5em $w_{t+1,n}\cm{x_{lose}} +=  \big(1-\Gamma(\bar{\alpha}_{t,n}, \alpha_{t,n}\cm{x}, T_{t,n}\cm{x})\big) w_{t,n}\cm{x} $\;
    }

    \For{$t = 0,\ldots, \tau$}
    {
        % $\bar{\alpha}_{t,n+1} = w_t\bar{\alpha}_{t,n} + \int_{\mathcal{R}}\alpha_{t,n}\cm{x}w_{t,n}\cm{x}dx$\;
        
        $\bar{\alpha}_{t+1,n+1} = \gamma \bar{\alpha}_{t,n} + \sum_x \alpha_{t,n}\cm{x} w_{t,n}\cm{x}$
    }
    
    \If{$\bar{\alpha}_{n+1} == \bar{\alpha}_n$}{
    break\;
    }

    }
     % \KwResult{$\alpha_{t,N}$}
    \caption{Compute $\alpha, T$ at equilibrium}
\end{algorithm*}

In this section, we detail the algorithm that we use to solve for equilibrium behavior of miners in our game. 

\subsection{Miner Wealth}
We model miner wealth at time $t$ using a density function $w_t$, where $w_t(x)$ is the fraction of miners with wealth $x$ at time $t$ (and $\sum_x w_t(x)=1$ for all $t$).

All miners begin with the same wealth, and a miner's actions are limited by their current wealth just as they would be in a real system. 
Recall that $c$ is the cost per unit of mining power, then the maximum $\alpha$ a miner with wealth $x$ can play is $x / c$, and any miner reaching wealth 0 will no longer be able to participate. 
For this reason, miners cannot simply optimize their expected reward at each time step; instead, they must also consider the rewards they might receive at future time steps as a function of their resulting wealth. 

\subsection{Value Function}
In order to optimize over future expected rewards, we introduce a value function into the miner's decision-making process. 
Slightly informally, the value function $V_t$ is defined as 

\begin{equation}
\begin{split}
    V_{t}(x) &= R(\bar{\alpha},\alpha,T) + \\ &\sum_{x'}  V_{t+1}(x') Pr[\text{wealth at }t+1=x'| \text{wealth at }t=x],
\end{split}
\end{equation}

where $Pr[\text{wealth at }t+1=x' | \text{wealth at }t=x]$ is the probability that a miner has wealth $x'$ at time $t+1$ given that they have wealth $x$ at time $t$.
Essentially, the value function describes the total reward a miner expects to gain over the remainder of the game as a function of their current wealth. 

\subsection{Approximating the Adversary's Decision Function}
In the model we have defined, there is a discontinuity in the mining reward function which arises from the adversary's `attack' or `do not attack' decision. 
For the sake of computational efficiency, we approximate $A(T, \bar{\alpha})$ using the sigmoid

\begin{equation}
    A'(T, \bar{\alpha}) = \frac{1}{1 + e^{-R_{adv}}}.
\end{equation}
This approximation is computationally convenient and does not materially impact our results.

\subsection{Computing the Equilibrium Policies}
Algorithm \ref{alg:Evaluation_alpha} solves for the optimal equilibrium policies for miners of varying wealth. 
It consists of three main computational steps which are executed in a loop until $\bar{\alpha}$ converges, meaning an equilibrium is reached.

\subsubsection{Compute optimal actions and associated value functions (lines 2-7)} 
We know that the value function at time $\tau + 1$ is 0 because the game concludes after $\tau$ rounds. 
This allows for the use of backwards recursion to solve for the optimal actions and value functions at each prior time step. 

Starting at time $t=\tau$, we compute the optimal $\alpha^*(x)$ and $T^*(x)$ as a function of wealth $x$.
Using these actions, we can compute the expected reward a miner receives at time $t$, their possible next wealth ($x_{win}$ or $x_{lose}$, depending on whether they win or lose the block) at time $t+1$, and the probability with which each next wealth is realized. 
This allows us to compute $V_t(x)$ as the summation of $R(\bar{\alpha}, \alpha^*(x),T^*(x))$, $V_{t+1}(x_{win})$, and $V_{t+1}(x_{lose})$ weighted by the probability of winning or losing the block.
We repeat this until reaching $t=0$.

\subsubsection{Compute wealth distribution (lines 8-13)}
At $t=0$ the wealth distribution is known. 
At each time step, we compute the probability that a miner with wealth $x$ wins or loses the next block when playing their best actions, $\alpha_{t,n}\cm{x}$ and $ T_{t,n}\cm{x}$.
For all $x$, the wealth distribution at time $t+1$ at wealth $x_{win}$ and $x_{lose}$ is incremented by the fraction of miners with wealth $x$ multiplied by their probability of winning and losing, respectively.

\subsubsection{Update $\bar{\alpha}$ (lines 14-16)}
The final step  is to compute the new values of $\bar{\alpha}_t$ for each time step.
These values are used in the next iteration of the outer loop until the algorithm converges. 
Fortunately, $\bar{\alpha}_t$ is simple to compute, as it is just the average of each $\alpha$ played weighted by the fraction of miners who played it.
Similar to that used in \cite{li2019mean}, we make use of a momentum parameter $\gamma \in [0,1)$ which slows the rate at which $\bar{\alpha}_t$ changes in order to prevent oscillatory behavior between iterations. 

An equilibrium is reached when $\bar{\alpha}_t$ converges for all $t$ (lines 17-19).

\section{Results}
\label{sec:perf}
We examine the behavior and performance (in terms of wealth gained) of miners in the presence of adversaries of varying strength. 
We show that the miners using $R$ as a reward function are far more profitable when a powerful adversary is present than those using $R_{naive}$. 

% In all figures, unless otherwise noted, we make use of the parameters specified in Table \ref{tab:params}, and all figures depict behaviors at equilibrium. 

% \begin{table}[h!]
%   \begin{center}
%     \caption{System Parameters}
%     \label{tab:params}
%     \begin{tabular}{c|l|c} % <-- Alignments: 1st column left, 2nd middle and 3rd right, with vertical lines in between
%       \textbf{Parameter} & \textbf{Description} & \textbf{Value}\\
%       \hline
%       $k$ & Confirmation Blocks & $6$ \\
%       $\beta$ & Adversary Fraction of Mining Power & .4\\
%       $b$ & Block Reward & 15\\
%       $c$ & Mining Cost & 1\\
%       $T_{max}$ & Maximum Transaction Value & 1000\\
%       $\tau$ & Time Steps & 5\\
%       $\gamma$ & Momentum & .9\\
%       $\lambda$ & Transaction Fee Constant & $.01$\\
%     \end{tabular}
%   \end{center}
% \end{table}

\subsection{Miners Using $R_{naive}$}
\begin{figure}[htp]
    \centering
    \includegraphics[width=\columnwidth]{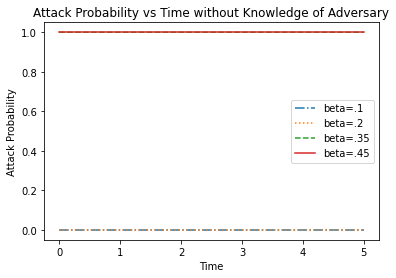}
    \caption{The adversary attack probability over time for different $\beta$ values. Note that the lines for $\beta=0.1$ and $\beta=0.2$ are overlapping, as are the lines for $\beta=0.35$ and $\beta=0.45$.}
    \label{fig:avtp}
\end{figure}

We first examine the scenario in which an adversary is ready to attack, but miners do not account for the adversary in their reward function (i.e., they choose actions based on $R_{naive}$).
Recall that miners using $R_{naive}$ always select a transaction value of $T_{max}$ in order to maximize the transaction fees they might receive upon winning the block. 
Under these conditions, Figure \ref{fig:avtp} shows the probability with which the adversary attacks at each time step\footnote{Note that $\beta=0$ is excluded from this figure, as an adversary with no mining power is not able to attack.}.
When the adversary is weak, their probability of success is too low for the attack to be profitable under any of the experiment parameters we tested. 
However, for $\beta=.35$ and $\beta=.45$, the adversary attacks at all time steps because miners select $\alpha$ and $T$ without consideration for adversarial attack.

\begin{figure}[htp]
    \centering
    \includegraphics[width=\columnwidth]{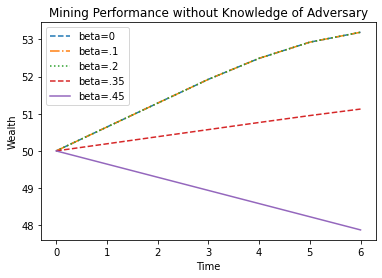}
    \caption{Evolution of average wealth over time for miners without knowledge of the adversary. Each line depicts mining performance of honest miners in the presence of adversaries of differing strength. Note that the lines for $\beta=0, \beta=0.1, \beta=0.2$ are identical and therefore overlap one another.}
    \label{fig:wealthvtimeperceived}
\end{figure}

Figure \ref{fig:wealthvtimeperceived} shows the result: for high values of $\beta$, mining performance is dramatically impacted, whereas for low values of $\beta$, mining performance is completely unaffected. 
Miners still remain profitable (albeit less so) against an adversary with 35\% of the mining power, but  mining performance continues to degrade as $\beta$ increases, and at $\beta=0.45$ miners are actually losing wealth in each time step.

\subsection{Miners with Knowledge of the Adversary}
\begin{figure}[htp]
    \centering
    \includegraphics[width=\columnwidth]{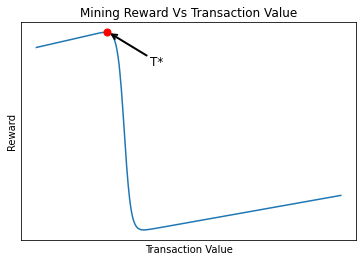}
    \caption{The honest mining reward as a function of transaction value selected when $\beta=0.4$. The point at which mining reward is maximized is $T^\star$, and the reward decreases sharply if a transaction value higher than $T^\star$ is selected, because in that regime it is profitable for the adversary to attack. Note that according to our model, there should be a discontinuity in the plot at $T^\star$. See Section \ref{sec:implementation} for details on why we smooth out this discontinuity in our implementation.}
    \label{fig:mvt}
\end{figure}

Next we examine the scenario in which an adversary is waiting to attack and miners \textit{do} account for it in their reward function. 
In this case, instead of selecting $T_{max}$, miners select a transaction value $T^\star$ which corresponds to the highest transaction value for which the adversary has no incentive to attack. 
Figure \ref{fig:mvt} can be used to visualize the shape of the expected reward for honest miners as a function of the transaction value they select.

\begin{figure}[htp]
    \centering
    \includegraphics[width=\columnwidth]{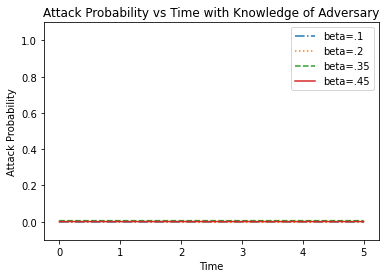}
    \caption{The adversary attack probability over time for different $\beta$ values. When the honest miners defend against attack, it is never profitable for the adversary to do so.}
    \label{fig:avta}
\end{figure}

Under the same experimental parameters as those used in the previous section, Figure \ref{fig:avta} shows that the adversary never attacks, as there is no time step during which the attack would be profitable. 

\begin{figure}[htp]
    \centering
    \includegraphics[width=\columnwidth]{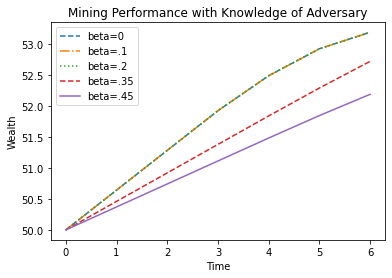}
    \caption{Evolution of average wealth over time for miners with knowledge of the adversary. Each line depicts mining performance of honest miners in the presence of adversaries of differing strength. Note that the lines for $\beta=0, \beta=0.1, \beta=0.2$ are identical and therefore overlap one another.}
    \label{fig:wealthvtimeactual}
\end{figure}

Figure \ref{fig:wealthvtimeactual} shows the performance of miners in this scenario. 
Unsurprisingly, the performance is identical for values of $\beta \leq 0.2$, as the adversary is simply too weak to affect mining performance (i.e., $T^\star \geq T_{max}$). 
However, for higher values of $\beta$, miners using $R$ as a reward function are able to effectively counter the adversary. 
In this scenario, miners achieve positive wealth in each time step, albeit with a slight reduction in performance resulting from the lower transaction fees collected due to the choice of $T^\star < T_{max}$. 

\subsection{Mining Performance Comparison with No Adversary}
\begin{figure}[htp]
    \centering
    \includegraphics[width=\columnwidth]{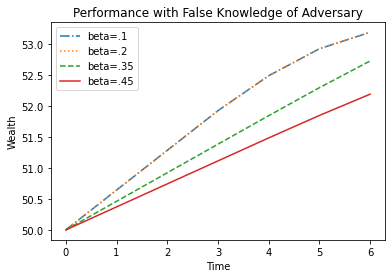}
    \caption{Evolution of average wealth over time for miners selecting their actions under the presumed threat of attack when no adversary is actually present. In this figure, $\beta$ values correspond to the adversary strength the miners are defending against, and the (overlapping) plots for $\beta=.1$ and $\beta=.2$ can be used as a baseline to determine mining performance without consideration for the adversary. }
    \label{fig:no_adv}
\end{figure}

There is an inherent tradeoff between using $R$ and $R_{naive}$, as miners must deliberately accept lower transaction fees in order to remove the incentive for the adversary to attack.
In order to understand this tradeoff, we also consider the case that the honest miners optimize against the threat of an adversarial attack but no adversary is present. 
Figure \ref{fig:no_adv} shows the result of this experiment. 

% As a benchmark, we also plot the results of an experiment in which miners optimize against $R_{naive}$ \textit{and} there is no adversarial attack--this line is marked ``No adv" and represents the best case scenario in which miners are not aware of an adversary and no adversary presents.
Interestingly, Figure \ref{fig:no_adv} is identical to Figure \ref{fig:avta} despite the fact that these experiments were conducted with adversaries of differing strength. 
This is expected, as the purpose of optimizing against $R$ is to remove the ability for the adversary to profit from the attack. 
From the perspective of the miners, there is no difference between the scenario in which an adversary could attack but does not and that in which there is no adversary. 

The takeaway from Figure \ref{fig:no_adv} is that, in the absence of an adversary, mining can be profitable for miners using $R$ as a reward function, although less so than for those using $R_{naive}$.
The degree to which performance is impacted depends on specific parameters of the system, and we note that our model does not capture the broader incentive miners might have to prevent attacks from being perpetrated against the system. 
We leave a more comprehensive characterization of this tradeoff to future work. 

% \section{Wealth Distribution vs Time}

% \begin{figure}[htp]
%     \centering
%     \includegraphics[width=\columnwidth]{figures/wealth_evolution.png}
%     \caption{Evolution of average wealth vs time for different block rewards when miners use the naive reward function vs the actual reward function.}
%     \label{fig:wealth_evolution}
% \end{figure}

% \section{Varying Block Rewards}

% \begin{figure*}[htp]
%     \centering
%     \includegraphics[width=16cm]{figures/block_reward.png}
%     \caption{Evolution of average wealth vs time for different block rewards when miners use the naive reward function (red) vs the actual reward function (blue).}
%     \label{fig:block_reward}
% \end{figure*}

\section{Analyzing the Bitcoin Network}
\label{sec:bitcoin}
In this section we analyze the Bitcoin protocol in order to determine its resiliency to double-spend attacks under our model. 
We find that, under the assumption of a 6 block confirmation time, the actual value throughput of the Bitcoin network is secure against an adversary controlling at most 25\% of the network mining power. 

\subsection{Estimating Bitcoin Network Parameters}
We pulled Bitcoin network statistics from Blockchain.com \cite{blockchaintxfees, blockchaintxvolume} between the dates of April 10, 2023 to April 17, 2023. 
During this time, the average value transferred by each block was 774.84 BTC, the block reward was 6.25 BTC, and the average total transaction fees per block were 0.16 BTC.

While the contents of a Bitcoin block are public knowledge, the mining costs paid by each miner are not. 
We avoid complex and inevitably inaccurate estimations of the purchase price of computational hardware amortized over its lifespan, varying electricity costs across time and geography, and other assorted operational expenses that would be required to calculate the cost to mine.  
Instead, we only assume that miners are profitable on average, meaning the total cost to mine each block during the time interval under study was at most 6.41 BTC\footnote{While this may be an overestimate, as miners likely would not mine for 0 profit, it does provide a rough upper bound on the cost to mine (and therefore the cost to attack).}. 

\subsection{Security Against Double-Spend Attacks}
Using these parameters, we solve for the maximum transaction value $T^\star$ that can be safely transacted in each block under various adversarial conditions. 
Recall that $T^\star$ is the transaction value for which the adversary's expected reward is 0. 
Then using Equation \ref{eq:adv_reward} we have 

\begin{equation}
    T^\star + f(T^\star) = \frac{C(\bar{\alpha}, \beta)}{P(\beta)} -  (k+1)b. 
\end{equation}

\begin{figure}[htp]
    \centering
    \includegraphics[width=\columnwidth]{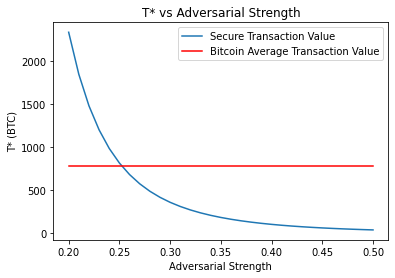}
    \caption{Under the estimated Bitcoin network parameters, the blue curve shows the amount of transaction value which can be transferred within a Bitcoin block for various adversarial strengths. The red line shows the actual amount of transaction value transferred in each Bitcoin block on average over the time interval we consider. According to this analysis, an adversary with 26\% of the network mining power would profit (in expectation) from a double-spend attack.}
    \label{fig:btc_plot}
\end{figure}

Under Bitcoin's constant fee function of $f(T^\star) = 0.16$, a required number of block confirmations $k=6$, and a total network mining cost of 6.41 BTC per block, we plot $T^\star$ with respect to $\beta$ in Figure \ref{fig:btc_plot}.
We find that an adversary controlling less than 25\% of the network mining power expects to lose money by attacking Bitcoin. 
As the attacker exceeds 25\% of the mining power, however, the double-spend attack becomes increasingly profitable. 

\subsection{Simulating a Percentage-Based Transaction Fee}
\begin{figure}[htp]
    \centering
    \includegraphics[width=\columnwidth]{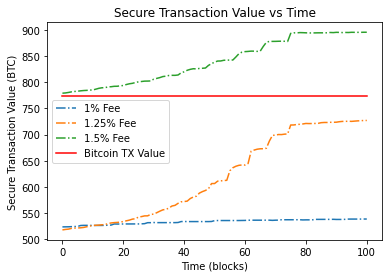}
    \caption{Using the Bitcoin network parameters we estimate, this plot shows $T^\star$ in the presence of a $\beta = 0.3$ adversary for several values of $\lambda$. We find that a percentage-based fee of between 1.25\% and 1.5\% would secure the actual value throughput that Bitcoin experienced during the time-frame under study. }
    \label{fig:val_vs_time}
\end{figure}

While the current network mining power seems insufficient to support the value throughput Bitcoin experiences, imposing a fixed value limitation on each block is not a satisfactory solution. 
Instead, the value limitation should grow or shrink with the needs of the network. 
In order to demonstrate this, we solve for the optimal strategies of Bitcoin miners who receive a 1\% transaction fee rather than a constant transaction fee.
In this simulation, miners always select $T^\star$, meaning there is no time at which the adversary can profit from an attack.
As additional transaction fees are collected, miners are able to increase the amount of mining power they contribute to each block, thereby increasing the amount of value which can be safely transferred. 
Figure \ref{fig:val_vs_time} shows $T^\star$ as a function of time in the presence of a $\beta = 0.3$ adversary under various fee functions. 
As expected, implementing a percentage-based fee function results in an increase in $T^\star$ over time, as the best strategy of each miner is to convert the extra wealth gained into additional mining power.
In turn, this additional mining power allows for a higher transaction value to be securely transferred by each block.

\section{Discussion and Future Work}
\label{sec:conclusions}

In this section, we  discuss the implications of this work and possible extensions of our model. 

\subsection{Implications for Bitcoin}
Our analysis in Section \ref{sec:bitcoin} does not imply that Bitcoin is immediately vulnerable to an attacker with 26\% of the network mining power, as there are many variables that our model does not account for.
Most notably, the entities receiving large transactions are (hopefully) waiting for many additional block confirmations beyond $k=6$ before declaring their transactions settled, and the attacker's probability of success decreases with each additional block confirmation.
A double-spend attack which is profitable when only 6 block confirmations are required may not be when 20 block confirmations are required. 

In other words, Bitcoin's resilience to double-spend attacks is dependent on the number of block confirmations required by the entities receiving large transactions.
If these entities were to require an insufficient number of block confirmations (whether it be through incorrect calculations or negligence), then Bitcoin would be vulnerable. 
Rather than Bitcoin users who receive large transactions, we argue that Bitcoin miners should operate the mechanisms which protect Bitcoin from attack. 
If miners adhere to a value limitation on each block, then no transaction will require additional block confirmations beyond the number specified by the protocol.

\subsection{The Adversary}
The adversary we present is somewhat limited in its strategy.
It can only decide whether or not to attack at each time step and has no ability to call off an attack early (if the honest miners pull too far ahead) or continue an attack for longer if they are not very far behind.

Additionally, we currently treat the adversary as being external to the system prior to launching an attack. 
This is a plausible modeling decision under the assumption that the adversary can quickly reallocate their mining power amongst different cryptocurrencies, either by renting computational power or swapping between different cryptocurrencies which use the same mining function, but it may not always hold in practice. 
Moreover, understanding the case where an ``honest" miner could opportunistically attempt a double-spend is also critical to the long-term security of these systems.

\subsection{Transaction Fees, Transaction Value, and Resistance to Attack}
There is an implicit relationship between the maximum transaction value that can be included in a block, the transaction fees a miner receives, and the anticipated mining power of an adversary. 
Higher value transactions increase the incentive for an adversary to attempt an attack, greater network mining power can increase the cost of attack, and higher transaction fees can be used to attract additional mining power to the network.
Understanding this relationship in greater detail would allow for the design of systems which secure a desired transaction value per block against rational attack without wasting resources on excess mining.
We believe this to be an exciting direction for future work.

% \subsection{Transaction Fees for Low-Value Transactions}
% In this work we make use of a transaction fee function which increases monotonically with transaction value. 
% While such a function makes sense in terms of security against double-spend attacks, it could result in very low-value transactions being ignored by miners in favor of transactions paying higher fees. 
% In practice, the transaction fee function could be implemented as a \textit{minimum} fee, while the actual fee paid could exceed this minimum. 
% % Or, more interestingly, a fee function could be developed which incorporates the \textit{quantity} of transactions being processed in addition to their value.

\subsection{Conclusions}
In this work, we model the mining game for Proof-of-Work cryptocurrencies in the presence of a rational, powerful adversary as a mean-field game. 
We model the reward functions of honest miners under threat of adversarial attack design and implement an algorithm that can efficiently solve for equilibrium policies of this game.
We show that miners using our adversary-aware reward function are able to eliminate the threat of attack from a profit-seeking adversary with only a small tradeoff in profitability. 

\bibliographystyle{IEEEtran}
\bibliography{References}

% \bibitem{b1} G. Eason, B. Noble, and I. N. Sneddon, ``On certain integrals of Lipschitz-Hankel type involving products of Bessel functions,'' Phil. Trans. Roy. Soc. London, vol. A247, pp. 529--551, April 1955.
% \bibitem{b2} J. Clerk Maxwell, A Treatise on Electricity and Magnetism, 3rd ed., vol. 2. Oxford: Clarendon, 1892, pp.68--73.
% \bibitem{b3} I. S. Jacobs and C. P. Bean, ``Fine particles, thin films and exchange anisotropy,'' in Magnetism, vol. III, G. T. Rado and H. Suhl, Eds. New York: Academic, 1963, pp. 271--350.
% \bibitem{b4} K. Elissa, ``Title of paper if known,'' unpublished.
% \bibitem{b5} R. Nicole, ``Title of paper with only first word capitalized,'' J. Name Stand. Abbrev., in press.
% \bibitem{b6} Y. Yorozu, M. Hirano, K. Oka, and Y. Tagawa, ``Electron spectroscopy studies on magneto-optical media and plastic substrate interface,'' IEEE Transl. J. Magn. Japan, vol. 2, pp. 740--741, August 1987 [Digests 9th Annual Conf. Magnetics Japan, p. 301, 1982].
% \bibitem{b7} M. Young, The Technical Writer's Handbook. Mill Valley, CA: University Science, 1989.
% \end{thebibliography}
% \vspace{12pt}
% \color{red}
% IEEE conference templates contain guidance text for composing and formatting conference papers. Please ensure that all template text is removed from your conference paper prior to submission to the conference. Failure to remove the template text from your paper may result in your paper not being published.

\end{document}